\pdfoutput=1
\documentclass[cits,hyper]{JINST}

\usepackage{booktabs}
\usepackage{setspace}

\title{First results on light readout from the 1-ton ArDM liquid argon detector for dark matter searches}
\author{
C.~Amsler$^a$, A.~Badertscher$^b$, V.~Boccone$^a$, A.~Bueno$^c$, M.~C.~Carmona-Benitez$^c$, 
W.~Creus$^a$, A.~Curioni$^b$, M.~Daniel$^d$, E.~J.~Dawe$^e$, U.~Degunda$^b$, A.~Gendotti$^b$, 
L.~Epprecht$^b$, S.~Horikawa$^b$, L.~Kaufmann$^b$, L.~Knecht$^b$, M.~Laffranchi$^b$, C.~Lazzaro$^b$, 
P.~K.~Lightfoot$^e$, D.~Lussi$^b$, J.~Lozano$^c$, A.~Marchionni$^b$, K.~Mavrokoridis$^e$ , 
A.~Melgarejo$^c$, P.~Mijakowski$^f $, G.~Natterer$^b$, S.~Navas-Concha$^c$, P.~Otyugova$^a$,  
M.~de~Prado$^d$, P.~Przewlocki$^f $, C.~Regenfus$^a$,
F.~Resnati$^b$, M.~Robinson$^e$, J.~Rochet$^a$, L.~Romero$^d$, E.~Rondio$^f$, A.~Rubbia$^b$, L.~Scotto-Lavina$^a$, N.~J.~C.~Spooner$^e$, T.~Strauss$^b$, J.~Ulbricht$^b$, and T.~Viant$^b$ 
(The~ArDM~Collaboration)\\
\llap{$^a$}Physik-Institut, University of Z\"urich, Winterthurerstrasse 190, CH-8057 Z\"urich, Switzerland\\
\llap{$^b$}ETH Zurich, Institute for Particle Physics, CH-8093 Z\"urich, Switzerland\\
\llap{$^c$}University of Granada, Dpto. de F\'isica Te\'orica y del Cosmos \& C.A.F.P.E,  Campus
Fuente Nueva, 18071 Granada, Spain\\
\llap{$^d$}CIEMAT, Div. de Fisica de Particulas, Avda. Complutense, 22, E-28040, Madrid, Spain\\
\llap{$^e$}University of Sheffield, Department of Physics and Astronomy,  Hicks Building,
Hounsfield Road, Sheffield, S3 7RH, UK \\
\llap{$^f$} The Andrzej Soltan Institute for Nuclear Studies, \\Ho\.za 69, 00-681 Warsaw, Poland
}

\abstract{ArDM-1t is the prototype for a next generation WIMP detector measuring both the scintillation light and the ionization 
charge from nuclear recoils in a 1-ton liquid argon target.  The goal is to reach a minimum recoil energy of 
30\,keVr to detect recoiling nuclei. In this paper we describe the experimental concept and  present results on 
the light detection system, tested for the first time in ArDM on the surface at CERN. With a preliminary and incomplete
set of PMTs, the light yield at zero electric field is found to be between 0.3-0.5~phe/keVee depending on the position within
the detector volume, confirming our expectations based on smaller detector setups.}


\begin{document}


\section{Introduction}

Astronomical observations give strong evidence for the existence of non-luminous and non-baryonic 
Dark Matter, presumably composed of a new type of elementary particle. It is of utmost importance to experimentally
verify if Dark Matter is indeed composed of the leading candidate called the Weakly Interacting Massive Particle (WIMP) \cite{Steigman}. 
If such WIMPs exist and are sufficiently stable, they would form a cold thermal relic gas in the present universe. Direct 
detection is  achieved by observing the 
energy deposited when they scatter elastically from  target nuclei, and with a cross-section magnitude which is 
predicted to be weak-like. 
This requires the capability 
of measuring a few events with recoil energies in the region of a few tens of keV and with negligible backgrounds from other
natural sources of radiation. Assuming a ``canonical'' WIMP halo model
and a mass of 100~GeV, a WIMP-proton cross-section of $10^{-44}$~cm$^2=10^{-8}$~pb would yield
about 1 recoil event per day per ton of argon above 30~keVr.
The direct laboratory detection of  WIMPs  is a necessary complement to e.g. SUSY searches actively pursued at  the 
LHC. 


The Argon Dark Matter 1-ton detector (ArDM-1t) is a   liquid argon (LAr) double-phase time projection chamber for direct 
dark matter searches \cite{ArDM}. The goal is to design, assemble and operate 
a  ton-scale liquid argon detector with independent ionization and scintillation readouts -- complementary to other
single and double phase noble liquid gas experiments like DEAP-CLEAN~\cite{Boulay:2008zz} and WARP~\cite{warp}
using argon, and XENON\cite{Aprile:2010zz}, XMASS~\cite{Suzuki:2008zz} and ZEPLIN-III~\cite{Akimov:2006qw} using xenon -- 
and to demonstrate 
the feasibility of an argon-based  ton-scale experiment for WIMP detection, with good detection efficiency 
and high background discrimination. A 1-ton prototype is presently installed on the surface at CERN, and 
has been operated for the first time in single-phase light collection mode, while double-phase operation is 
in preparation. 

Substantial R\&D efforts  were needed to develop the light readout of ArDM-1t,
justifying the operation of the detector first on the surface, where many functionalities could more easily be tested, before moving to 
an underground location for physics runs in low background conditions. 
The developments that resulted in the definition of the ArDM reflector were reported in~\cite{ArDMWLS}. In this paper,
we present results obtained during the first operation of the light detection system fully embedded
in the ArDM-1t vessel.

\section{Detector}

\begin{figure}[htb]
\centering
\includegraphics[width=0.5\textwidth]{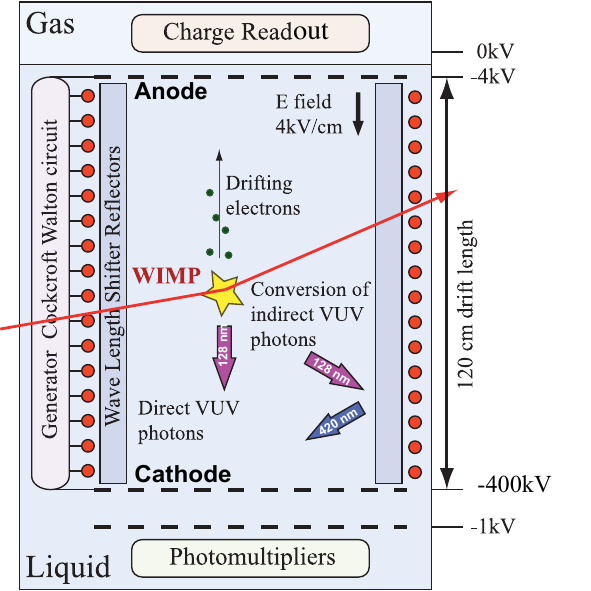}
\caption{Conceptual layout of the ArDM experiment.}
\label{fig:ArDMdetector1}
\end{figure}

\begin{figure}[htb]
\centering
\includegraphics[width=0.65\textwidth]{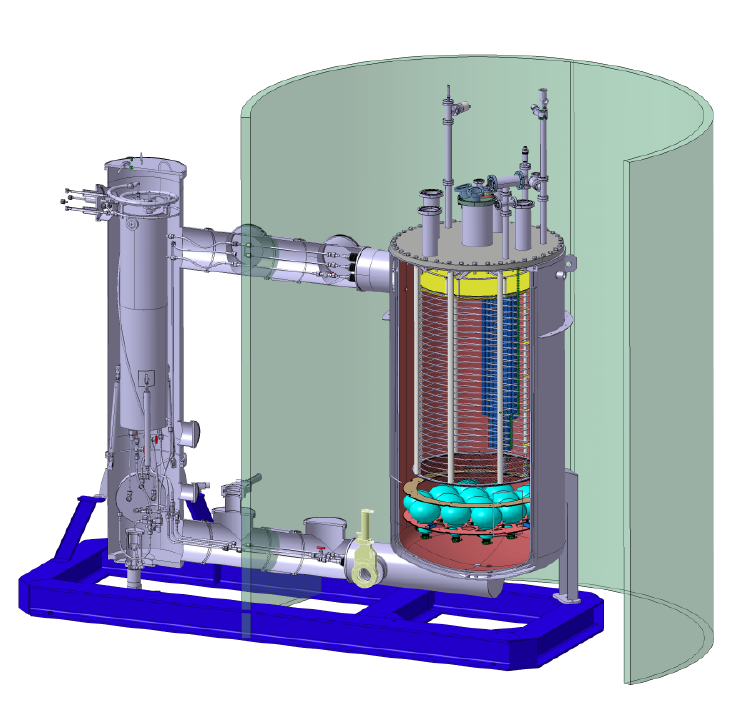}
\caption{3D model of the cryo-system, main dewar and detector components. The main vessel inner diameter is 100~cm and
the maximal drift length is 120~cm. Yellow: charge readout system, cyan : light readout PMTs,
green : outer dimension of neutron shield.}
\label{fig:ArDMdetector2}
\end{figure}

Figure \ref{fig:ArDMdetector1}  shows a conceptual layout of the apparatus.
The sensitive liquid argon volume diameter, delimited by the wavelength shifter reflectors, is 80~cm and the maximal drift length is 120~cm. 
%
 %
The ultra-clean liquid argon in the detector is provided and maintained 
by a cryogenic and purification system separated from the main dewar to 
allow  the insertion of a radiation shield against neutrons (See Figure~\ref{fig:ArDMdetector2}).
Charged particles lead to ionization and 
excitation of argon atoms, forming excimers. The luminescence is detected by an array of 
photomultipliers immersed in the liquid at the bottom of the detector, while the charge (electron cloud) is 
drifted in a strong electric field towards a charge amplifying readout system located in the vapor phase above the 
liquid argon. 

The basic mechanisms by which slow ionizing particles lead to ionization and scintillation signals in liquid argon 
in energy regions of tens of keV are not well known.
However, to define the needed performance of the detector, it is sufficient to assume that
a WIMP-induced nuclear recoil of 30 keVr will produce around 300  VUV photons (at 128~nm), 
together with a few free ionization electrons, the latter number depending on the electric field strength due to electron-ion recombination~\cite
{ChandrasekharanPhD}.  For background discrimination (e.g. from background $\gamma$'s), 30 keVr recoils should be compared to an electronic
energy deposition of 30~keV$\times L_{eff}\approx 7.5$~keVee\footnote{The ratio of the nuclear recoil scintillation response to the electronic recoil response was
recently measured to be $0.25 \pm 0.02 \pm 0.01(correlated)$ for recoils above 20~keVr. See Ref.~\cite{Gastler:2010sc}.}, which will generate several hundreds of free ionization electrons.
In addition, a fast (few ns) and a slow ($\sim$ 1.6 $\mu$s) component of the scintillation light are observed in liquid 
argon \cite{Hitachi}, their ratio also depending on the ionizing type.
Therefore, the charge/light ratio~\cite{Barabash:1989xr,IC1} and pulse shape discrimination~\cite{Boulay:2006mb,Lippincott:2008ad} provide efficient 
discrimination against gamma or beta electron recoils.
Heavily ionizing particles such as $\alpha$'s or nuclear recoils contribute mostly to 
the fast decaying component, while the contribution of electrons and $\gamma$'s  to the slow component 
is larger. Quantitatively, the component ratio $CR$ of the fast (< 50 ns) to the total intensity is typically 0.8 
for heavily  and 0.3  for minimum ionizing projectiles. Measurements at low energy 
show mean values within 0.6-0.7(nuclear recoils) and 0.36-0.28(electronic recoils) in the energy range 7-32~keVee~\cite{Lippincott:2008ad}.
The higher ratio of light to charge production for 
nuclear recoils, and the corresponding higher component ratio,  can both be used to reduce background 
in WIMP searches. 

A Greinacher (Cockroft-Walton) generator chain, as described in Ref.~\cite{Horikawa:2010bv}, is immersed in the liquid and
is able to provide a field up to about 3\,kV/cm 
to avoid electron-ion recombination and to drift the electrons to the top surface of the liquid towards the 
gas phase. At the interface, the electrons are extracted to the gas phase and accelerated towards 
the Large Electron Multiplier (LEM \cite{OtyugovaPhD, Badertscher:2009av,Badertscher:2008rf} or THGEM \cite
{THGEM1,THGEM2,THGEM3}), which provides multiplication and position reconstruction by a 
segmented anode. A 3D image of the event is thus  obtained with a much lower energy threshold than in 
a single phase liquid argon time projection chamber, allowing for example to detect multiple
nuclear recoils as expected from fast neutrons entering the detector. This information will be used to estimate
the irreducible single recoil neutron
background by direct extrapolation from the observed multiple scatter events, which constitute more than 50\% of the
neutron induced events within the large active volume of ArDM-1t~\cite{KaufmannPhD}.

The 128 nm VUV light is emitted isotropically from the interaction point and converted to blue (420\,nm) 
light by a wavelength shifter (WLS) deposited on reflectors mounted on the side walls which also define
the fiducial volume~\cite{ArDMWLS}. The reflectors consist of Tetratex (TTX) foils, 254\,$\mu$m thick, which have nearly 
100\% diffuse Lambertian reflectance. A surface density of 1\,mg/cm$^2$ of TetraPhenyl Butadiene (TPB) 
was evaporated on these foils \cite{Hugo}. In the completed detector, the shifted and reflected light is collected at the bottom of the 
cryostat by 14 photomultiplier tubes (PMT, 8" Hamamatsu  R5912-02MOD  with bialkali photocathodes 
and Pt-underlay) immersed in the liquid.  Each PMT is soldered with its leads onto a 3\,mm thick printed 
circuit board, providing also the individual mechanical footing and the voltages for cathode and dynodes with 
passive electronic components.  A support structure holding up the fourteen 8'' PMTs was built to sustain 
the total buoyant force in LAr ($\approx$\,60\,kg) under the constraint of minimal amount of material. It 
was manufactured out of  a 5\,mm thick stainless steel disk by cutting out hexagonal holes. 
The individual PMT units were fixed to this plate by 3 screws through their PCB bases 
on the bottom. 

A WLS deposition on the PMT glass (0.05\,mg/cm$^2$ of TPB) was also performed to improve the direct 
VUV light detection. To optimize  the  light readout, a range of reflectors and WLS depositing 
combinations were investigated with several small prototypes, in which argon scintillation light was 
generated by radioactive sources in gas at normal temperature and pressure \cite{Buchler}. A detailed 
comparison is reported in Refs.~\cite{ArDMWLS} and \cite{BocconePhD}.

\section{Performance of the light readout system}

After a several-month-long ultra-high vacuum evacuation of the main detector vessel, which removed
outgassing from detector surfaces and materials, certified leak-tightness and finally resulted in a
residual pressure $\ll 10^{-5}$~mbar in the full volume encompassing the detector and the purification circuit,
a four weeks engineering run of ArDM-1t with a preliminary version of the light readout
took place for the first time in spring 2009.  This first run
was preceded by a careful assessment of the cryogenic requirements  and the related safety issues.

\begin{figure}[htb]
\parbox{50mm}{\mbox{
\includegraphics[width=40mm]{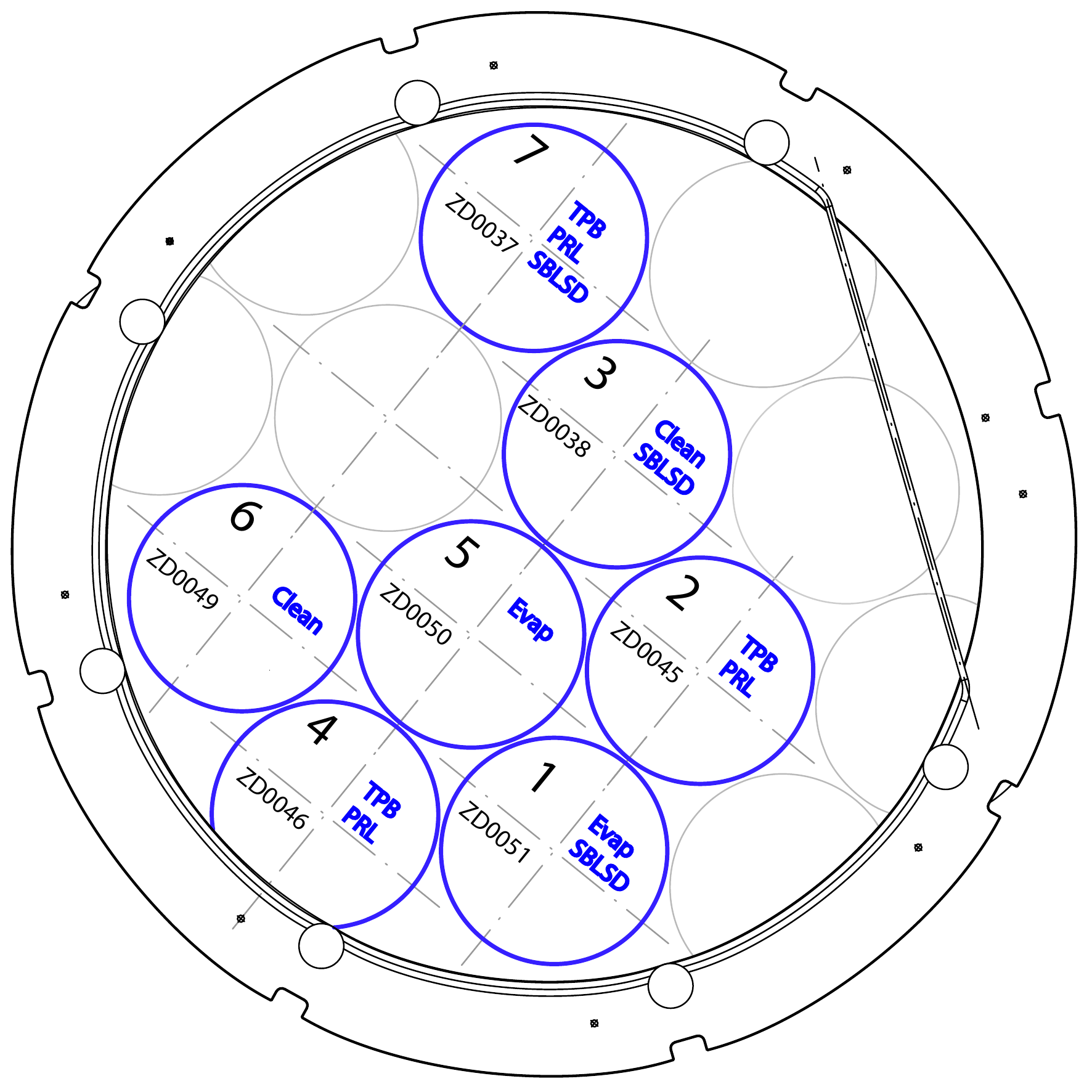}
}\centering}\hfill
\parbox{50mm}{\mbox{
\includegraphics[width=40mm]{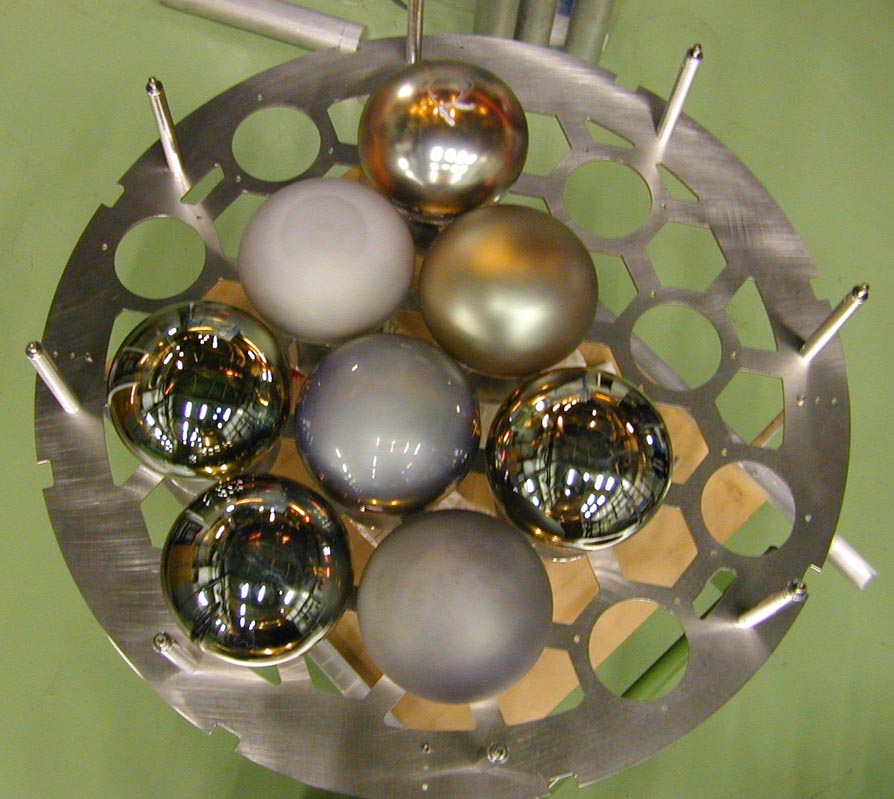}
}\centering}\hfill
\parbox{50mm}{\mbox{
\includegraphics[width=25mm]{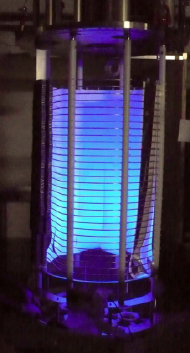}
}\centering} \caption[]{Left: schematic showing position of the temporary 7~PMTs as listed in Table 1. Middle: photograph showing the top view of 
the  temporary 7 PMT array  used for the first measurements in liquid argon. Right: photograph showing the TPB covered lateral TTX foils within 
the drift field shaper ring electrodes, illuminated with UV light.
\label{fig:PMT7}}
\end{figure}

Large amounts of calibration data with different radioactive sources were collected. 
During this first run a reduced temporary set of 7 PMTs with various surface treatments 
(sand-blasted or transparent glass, no PMT surface coating, TPB-Paraloid coating
or TPB-evaporation -- for details on the various methods see Ref.~\cite{ArDMWLS}) was installed to test  the efficiency for the direct light 
detection (an 8th PMT was installed but could not be operated due to HV problems)~\cite{Prelres}. 
The used PMT array and support 
structure are shown in  fig.\,\ref{fig:PMT7}, and specification details are given 
in Table\,\ref{table:PMT7}.
These measurements indicated a slightly better performance  (at the 10 \% level) from evaporating 
a thin layer of TPB on the PMT entrance windows instead of coating them with a thin TPB-Paraloid layer. 
The first process could also be done with better uniformity. More details on this as well as 
comparative measurements in gaseous argon can be found in \cite{BocconePhD}.
%
Figure \ref{fig:PMT7} (right)  shows the lateral WLS foils under UV illumination before insertion into 
the cryostat.

\begin{table}[hbt]
\begin{center}
\begin{tabular}{|c|cccll|}
\hline
\#  & type & dynodes & window & coating & max. gain \\ 
\hline \hline
1 &  Hamamatsu R5912-02MOD & 14  & Sandblasted & TPB evaporated & $\approx 1 \cdot 10^{9}$ \\
2 &  Hamamatsu R5912-02MOD & 14  & Transparent & TPB-Paraloid & $\approx 1 \cdot 10^{9}$\\ 
3 &  Hamamatsu R5912-MOD~~~~ & 10  & Sandblasted & No coating & $\approx 5 \cdot 10^{7}$\\
4 &  Hamamatsu R5912-02MOD & 14  & Transparent & TPB-Paraloid & $\approx 1 \cdot 10^{9}$\\
5 &  Hamamatsu R5912-02MOD & 14  & Transparent & TPB evaporated & $\approx 1 \cdot 10^{9}$\\
6 &  Hamamatsu R5912-02MOD & 14  & Transparent & No coating & $\approx 1 \cdot 10^{9}$\\
7 &  Hamamatsu R5912-MOD~~~~ & 10  & Sandblasted & TPB-Paraloid & $\approx 5 \cdot 10^{7}$\\
\hline
\end{tabular}
\caption{Properties of the cryogenic hemispherical 8'' PMTs installed in the LAr detector.
}
\label{table:PMT7}
\end{center}
\end{table}
\begin{figure}[hbt]
\begin{center} 
\includegraphics[height=5.2cm]{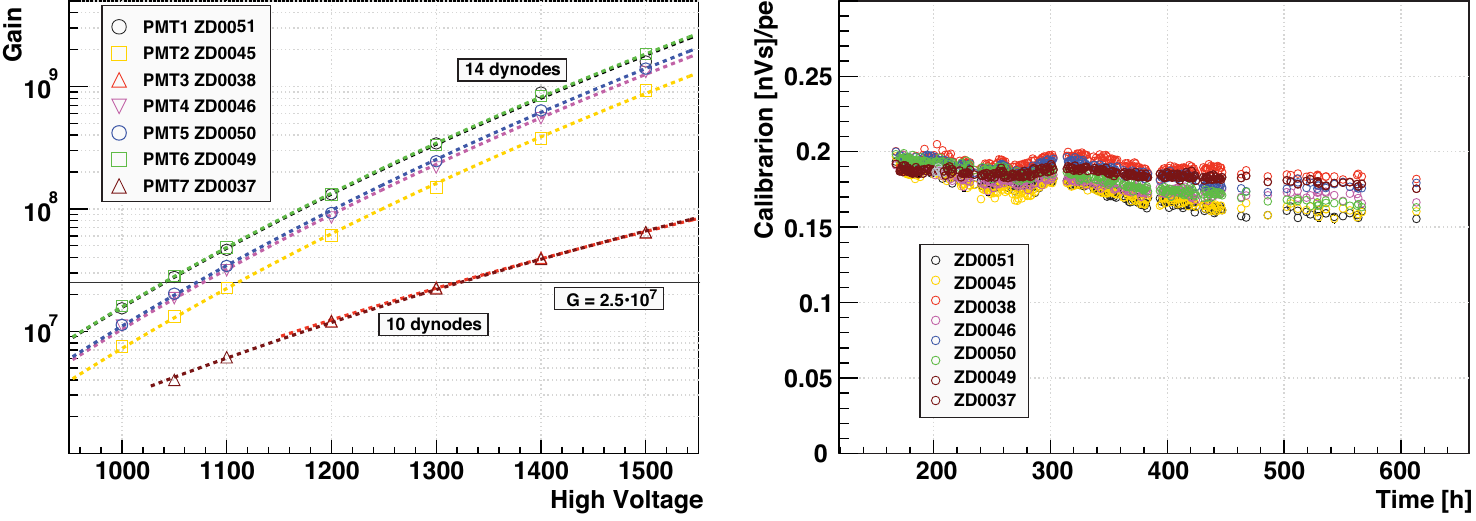}
\caption{Left: Gain curves of the PMTs in LAr. The curves are fits to the data points
(see text). Right: Time evolution of the calibration of the PMTs during the run in nVs/photoelectron (p.e.).}
\label{fig:ArDMGainCalibration}
\end{center}
\end{figure}

Each PMT was individually fed from one HV channel of a CAEN SY2527 system 
and read out by a 10\,bit FADC channel (1Gs/s) from an ACQIRIS DC 282 system.
The system could be triggered internally e.g. by requiring at least 2 PMTs with a signal over threshold, or 
externally, e.g. tagging an external radioactive source (see below). 

The gain of the PMTs was determined from the average single photon charge
observed from short light pulses generated by an internally mounted LED (400\,nm) 
placed in the vapor region above the liquid argon surface. 
Figure\,\ref{fig:ArDMGainCalibration} (left) shows the measured gain $G$
as a function of applied high voltage for the 7 temporary PMTs  immersed 
in LAr for several days. The data points were 
fitted with the function $G = A\cdot V^{kn}$, where $n$ is the number of dynodes and 
$A$, $k$ are fit parameters.
At the beginning of the run the PMT gains were set to $2.5\cdot 10^7$ by
adjusting individually the bias voltages. This value corresponds to an 
average integrated signal of 0.2\,nVs per photoelectron and was matching our
desired dynamic range. The number of photoelectrons  collected by each PMT 
was determined from the integrated charge measured over a  50~$\Omega$ resistor. 
Figure \ref{fig:ArDMGainCalibration}  (right) shows the time evolution of the calibration constants
during the full duration of the run. The small drift is attributed to slow temperature and pressure
fluctuations introducing mechanical stress on the dynode chain of the PMT, and appears to level off after some time 
in stable working conditions. It can be accounted for during the offline analysis.

\begin{figure}[hbt]
\begin{center} 
\includegraphics[height=6cm]{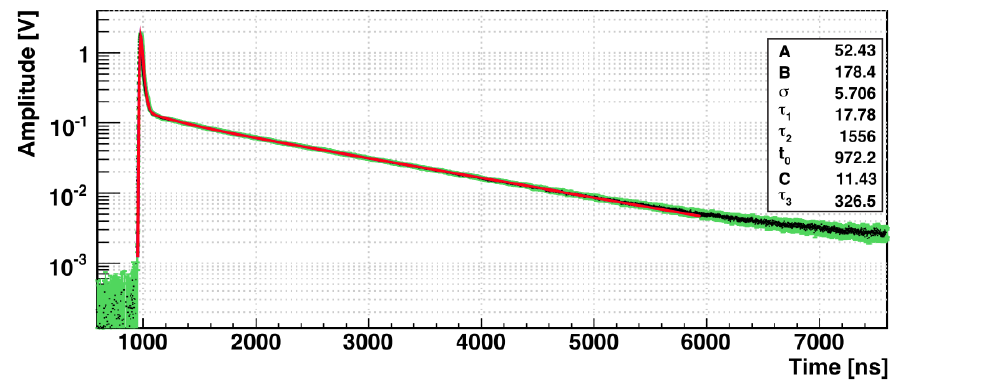}
\caption{Time distribution in LAr. The red curve is a fit with three exponential decay functions.}
\label{ArDMAvgPulseElectrons}
\end{center}
\end{figure}

Figure \ref{ArDMAvgPulseElectrons} shows a typical time distribution of the signal in the detector for electrons, $
\gamma$'s and cosmic muons which were selected by the component ratio $CR$ < 0.4. The data were taken with 
the internal trigger by setting the discriminator thresholds to  50 mV.
A fit with three exponential decay functions  leads to an excellent description of the observed average waveform data, with
measured lifetimes $\tau_1$ = 17.8 $\pm$ 1.0 ns for the fast component, $\tau_2$ = 1556 $\pm$ 10 ns for the slow 
component, and $\tau_3$ = 326 $\pm$ 10 ns for the intermediate one, with contributions of 21, 74 and 5\%, 
respectively \cite{BocconePhD}. 

\begin{figure}[hbt]
\begin{center} 
\includegraphics[width=0.8\textwidth]{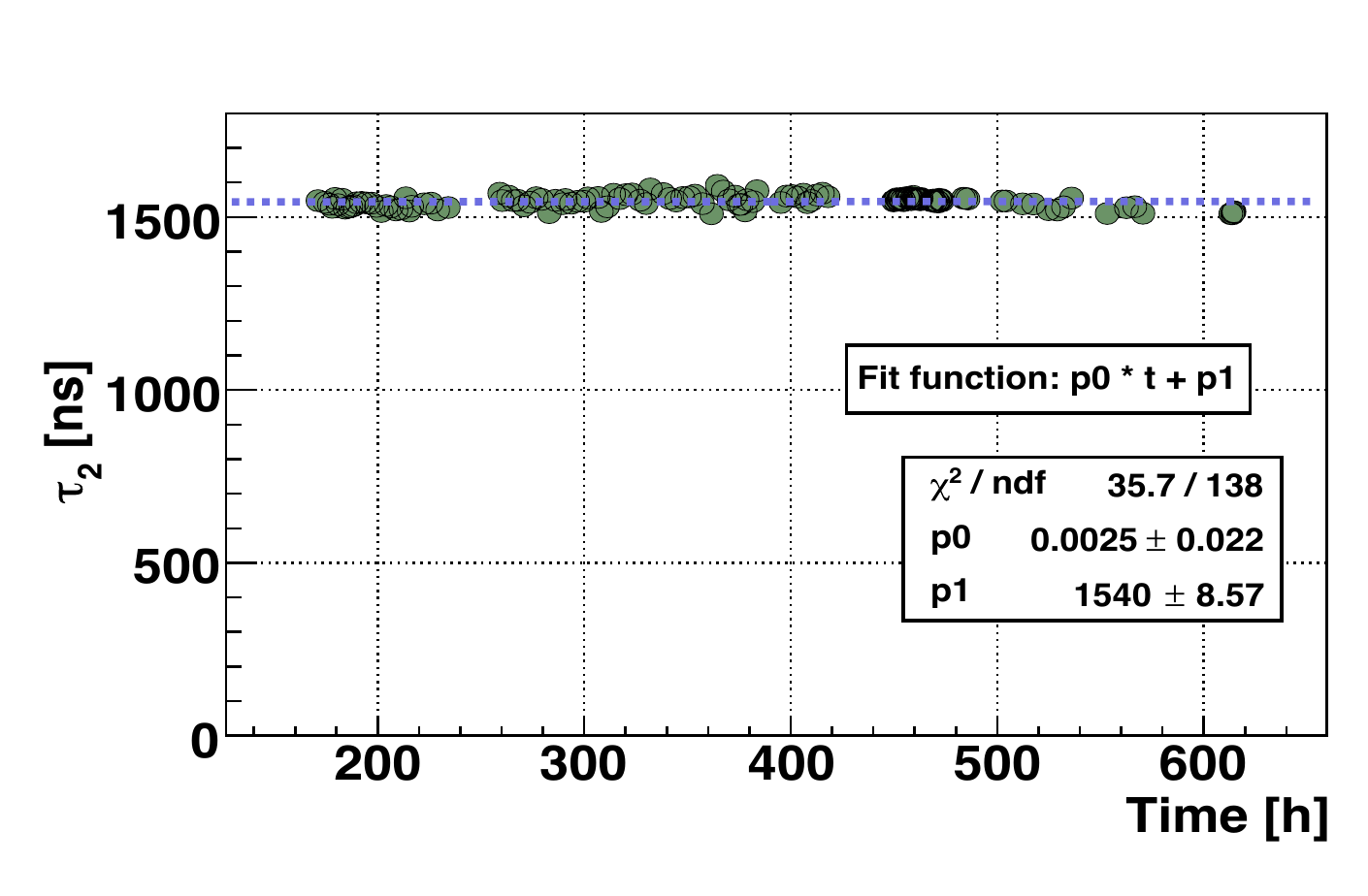}
\caption{Evolution of the lifetime $\tau_2$ of the slow scintillation component during the run. The dashed line is a straight line fit revealing good purity conditions.}
\label{fig:ArDMPurityStability}
\end{center}
\end{figure}

No attempt was made to deconvolute the response of the PMTs and the acquisition system
in $\tau_1$, while the fitted $\tau_2$ was used to qualify the achieved purity of the liquid argon in our setup, a very relevant quantity
for the performance of light and charge collection. 
Indeed, the mean lifetime $\tau_2$ of the slow scintillation component  is very sensitive to traces of impurities and can be 
used to monitor the LAr purity. This effect  was studied in gaseous argon (see ref. \cite{lumquench} for details) and 
similar observations are made in liquid \cite{BocconePhD, Acciarri}. Accordingly, the LAr purity was studied and monitored 
over several weeks by measuring  $\tau_2$ with radioactive sources. A single  exponential fit to the average pulse 
distribution at later times was applied since $\tau_2$ does not depend on the type of ionizing particle. 
Figure\,\ref{fig:ArDMPurityStability} 
shows the time dependence of $\tau_2$ during the run (600h). Our average-over-time result  (fit dashed line) $\tau_2$ = 1.540 $\pm$ 0.009 $\mu$s is 
in good agreement with expectations and consistent with no significant deterioration of purity 
over several weeks of operation.
From the observed lifetime of the slow component, we estimate that the concentration of N$_2$ and O$_2$ 
was consistent with being less than 0.1 ppm \cite{BocconePhD,Acciarri} Hence the detector was satisfactorily tight and  clean, 
although the active LAr recirculation system was not yet operated.

\begin{figure}[hbt]
\begin{center} 
\includegraphics[width=0.6\textwidth]{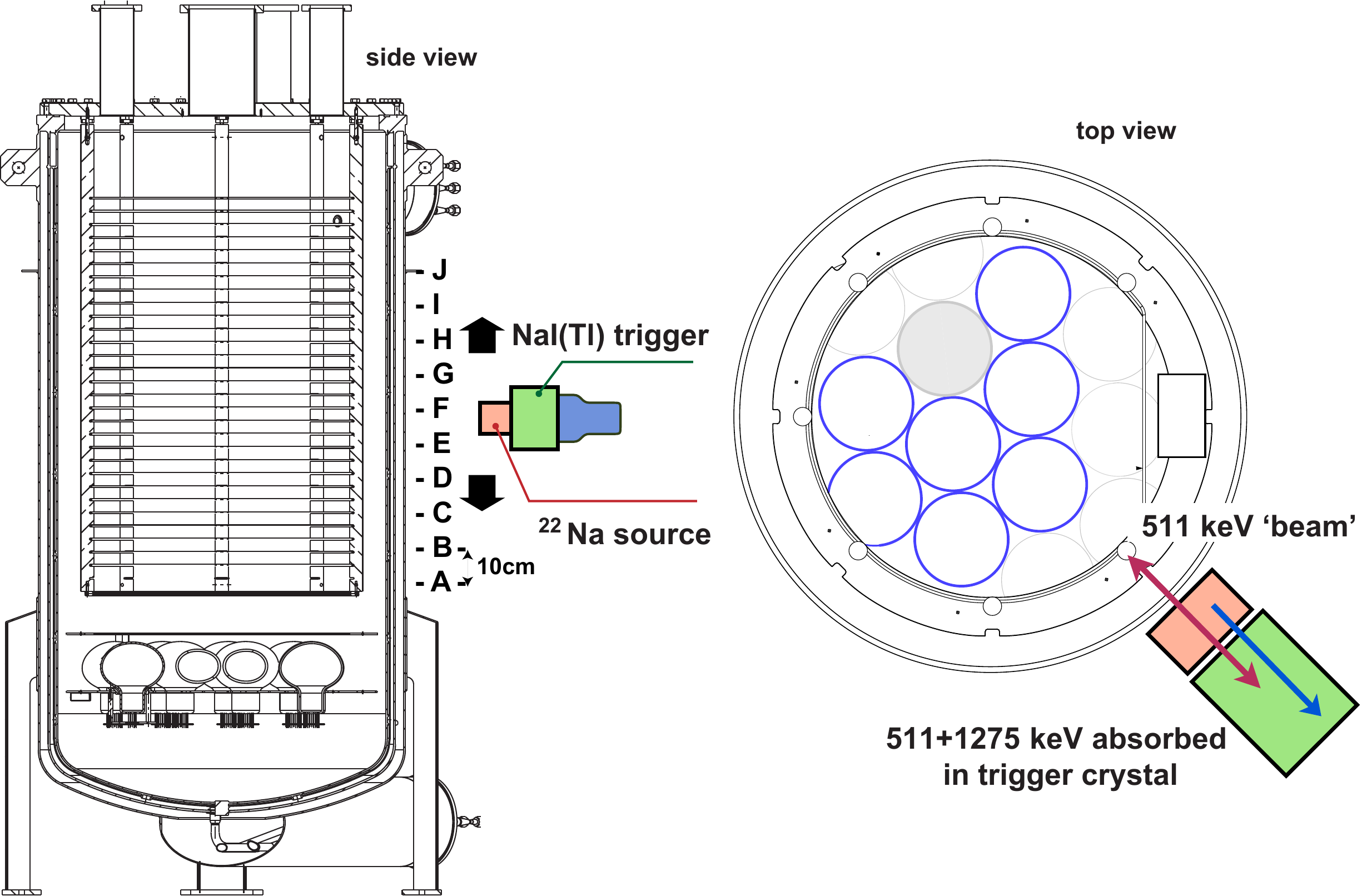}
\caption{Sketch of the apparatus used  to measure the light yield in liquid argon with an external $^{22}$Na $\gamma$-source and a triggering NaI(Tl) crystal.}
\label{fig:ArDMExternalSource}
\end{center}
\end{figure}

We now present our first evaluation of the detector performance with respect to light 
yield and detection efficiency of low energy events, which were the two main goals pursued 
during this first engineering run. The measurements were performed with external radioactive sources at zero electric 
field.  
For the measurements presented here a 20\,kBq $^{22}$Na $\gamma$-source was employed, 
delivering positrons (annihilating into two 511 keV $\gamma$'s) and monochromatic 1275\,keV $\gamma
$'s. 
The light yield produced by one of the 511\,keV $\gamma$'s, following (multiple) Compton scattering, was 
measured by triggering with a 4" NaI(Tl) scintillation crystal on the second 511\,keV emitted in the opposite 
direction, and on the 1275\,keV $\gamma$. 
A sketch of the experimental arrangement is shown in fig. \ref{fig:ArDMExternalSource}. 

The PMT array at the bottom of the apparatus is meant to measure the scintillation 
light with good time resolution. Single photon counting is possible with  a 
signal-to-noise ratio of around 20, the limiting noise being generated at the input stage of the
FADCs. The  coincidence of at least 2 PMT signals within typically ~10 -- 1000\,ns  was used as the event 
trigger and initiated the readout of 10000 FADC samples, corresponding to a 10\,$\mu$s long interval at 1 Gs/s with 
1000 (9000) samples before (after) the trigger time. 

Rather than summing the digitized samples to recover the PMT measured light intensity, which due to the slow component
of argon has a long tail of decaying amplitude, the time digitized 
waveforms were analysed offline to identify ``charge 
clusters'' coming
from groups of photons or from single photons. The
individual charge clusters were  then added in an optimized time window of 4500~ns located around the peak found with maximum amplitude,
to obtain integrated pulse heights. This method is very efficient in the case of the long decay time of argon, and was found to 
yield an integration error coming from electronic noise and accidental dark counts of about 1.4~p.e. (rms) at high energies, reaching the limit  of 1/20~p.e. for single 
photons. These values were determined from random empty events collected 10 $\mu$s before the signal trigger.
The pedestal of each PMT was determined on an event-by-event basis in an iterative way from the time 
distributions in the first 600\,ns before trigger time. 

\subsection{Monte Carlo simulation}

A full Monte Carlo (MC) simulation of the ArDM detector has been developed \,\cite{KaufmannPhD,CarmonaPhD}. The
code, based on GEANT4\,\cite{GEANT4}, includes tracking of all particles through a detailed detector geometry description
and has in addition a full simulation of the light propagation and detection in ArDM
(isotropic emission of the primary scintillation light, VUV propagation, wavelength shifting, 
surface optical properties, reflections, and PMT response). The following assumptions were made:
\begin{itemize}
\vspace{-1mm}\item The absorption of VUV or optical photons in LAr is neglected; Rayleigh scattering is included~\cite{Seidel};
\vspace{-2mm}\item The wavelength shifting efficiency of the side reflectors and of the PMTs coating (TPB) was 
100\%;
\vspace{-2mm}\item The diffuse reflection coefficient of the side reflectors was varied between 88\% and 98\%, with 
a value of 95\% giving the best description of the experimental data;
\vspace{-2mm}\item The maximum quantum efficiencies of the PMTs were  between 18.5\% and 22\%, according to 
the data sheets of each PMT, and the quantum efficiency over the photocathode surfaces were parameterized as a 
function of the polar angle, with a maximum on the vertical ($\theta=0$), decreasing by about 30\% for large $
\theta$;
\end{itemize}

These assumptions are in good agreement  with our own measurements on smaller prototypes  in the laboratory\,
\cite{ArDMWLS, BuenoJINST2008}. For instance, our measurements were 95\% for the diffuse reflection coefficient and typically 15\%  
for the quantum efficiency integrated over $\theta$. 
In addition, outside the GEANT4 MC:
\begin{itemize}
\vspace{-2mm}\item A Gaussian noise smearing of 1.4 p.e. was applied to MC data;
\vspace{-2mm}\item The measured values were corrected to compensate for our chosen finite integration time of 4500\,ns. 
\end{itemize}

These MC predictions, expressed in numbers of photoelectrons  
could then be compared directly with data.
For the absolute direct comparison with the external sources, the MC simulation was extended 
to include the test setup shown in fig. \ref{fig:ArDMExternalSource}, including the 
acceptance of the NaI(Tl) trigger counter. The $^{22}$Na external source was taken into account by simulating a 
511 keV $\gamma$ emitted in a cone covering the NaI solid angle.
Data taken with the $^{22}$Na-source at different heights were directly compared with MC simulated data
and could be absolutely normalized, since the detection efficiency of the 511~keV photon inside ArDM-1t was accordingly
properly included in the MC. The same selection cuts were applied to the data and MC.

\begin{figure}[hbt]
\begin{center} 
\includegraphics[width=0.95\textwidth]{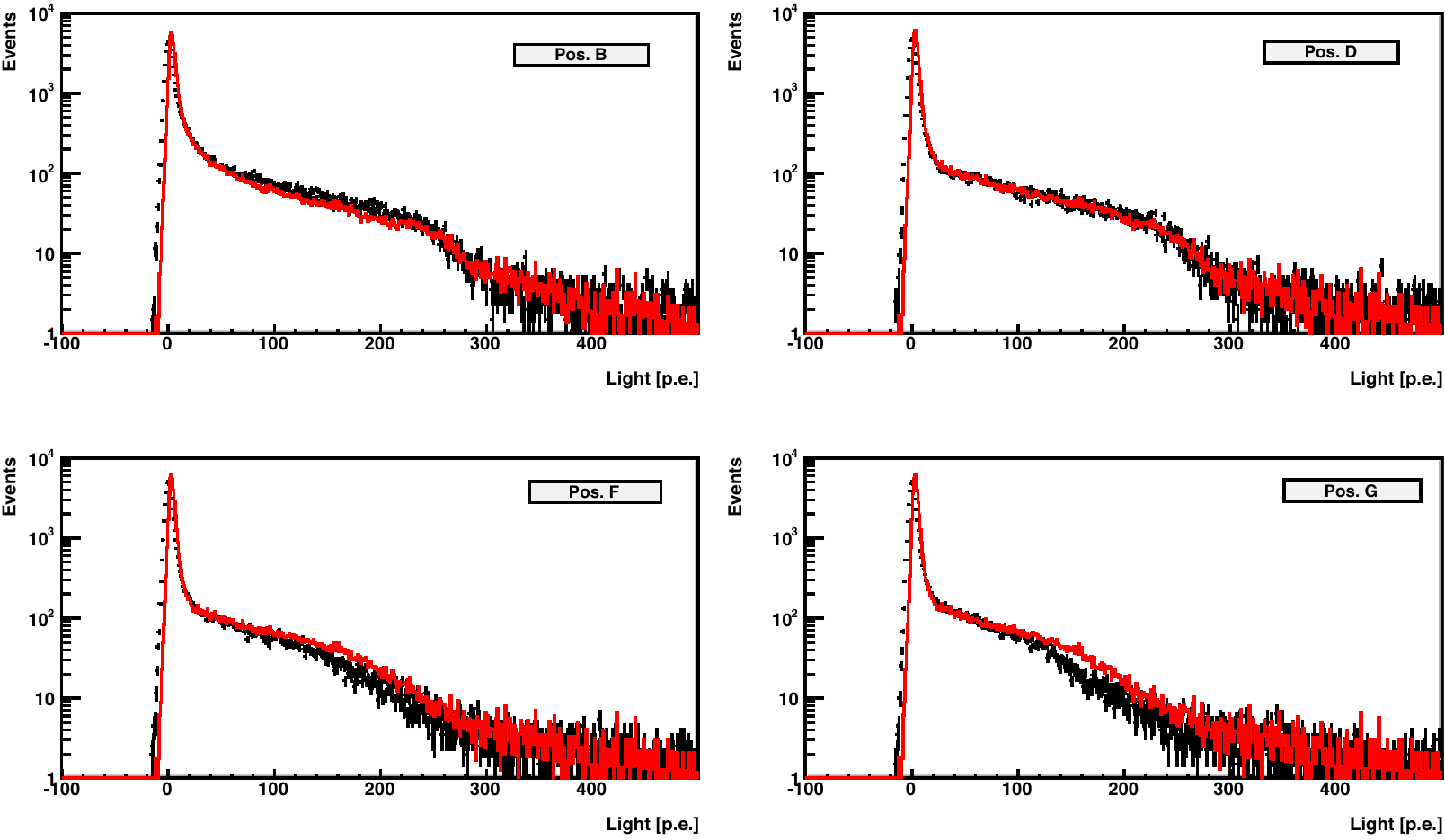}
\caption{Light detection  in the ArDM-1t detector for 511 keV $\gamma$'s, in photoelectrons at the positions B, D, F, G of the $^{22}$Na- source  (see fig. \protect\ref{fig:ArDMExternalSource}). The measurements are shown in black with error bars, the simulated data in red. The data were triggered by the coincident absorption of a 511 and a 1275 keV $\gamma$ in the external trigger NaI(Tl) crystal.}
\label{fig:DataMCComparison22Na}
\end{center}
\end{figure}

Figure\,\ref {fig:DataMCComparison22Na} shows the light yield distributions 
for the four positions (see fig.\,\ref{fig:ArDMExternalSource}) B, D, F and G (black bars), absolutely normalized to the MC. 
The light yield (in p.e.)  was also 
calculated by MC simulation (95\% reflectivity) as a function of  energy deposit, and the distributions (red line) compared to the 
measured ones. Good agreement is found, with discrepancies of less than 10\% in the light yield.  The drops in the spectra around 
200 -- 300 p.e. correspond to  full absorption of the 511\,keV photon in the fiducial volume. No clear full energy peak is visible due to 
the fact the most of the  511\,keV photons undergo multiple interactions and are not fully absorbed. The width of the peak at zero photoelectrons is 
determined from events collected with a special trigger setup, acquiring events happening 10 $\mu$s before the actual trigger, i.e. 
empty events, to take into account dark counts and accidental pileup.

A general trend of the spectra at the various source positions is the lower  light yield for events occurring in the 
upper part of the detector, which is well described by the MC simulation (however, we expect this effect to be less 
pronounced in the completed detector which will include the charge readout, due to downward reflection of the wavelength shifted light
from the reflective charge readout surface).

\begin{figure}[hbt]
\begin{center}
\includegraphics[width=0.5\textwidth]{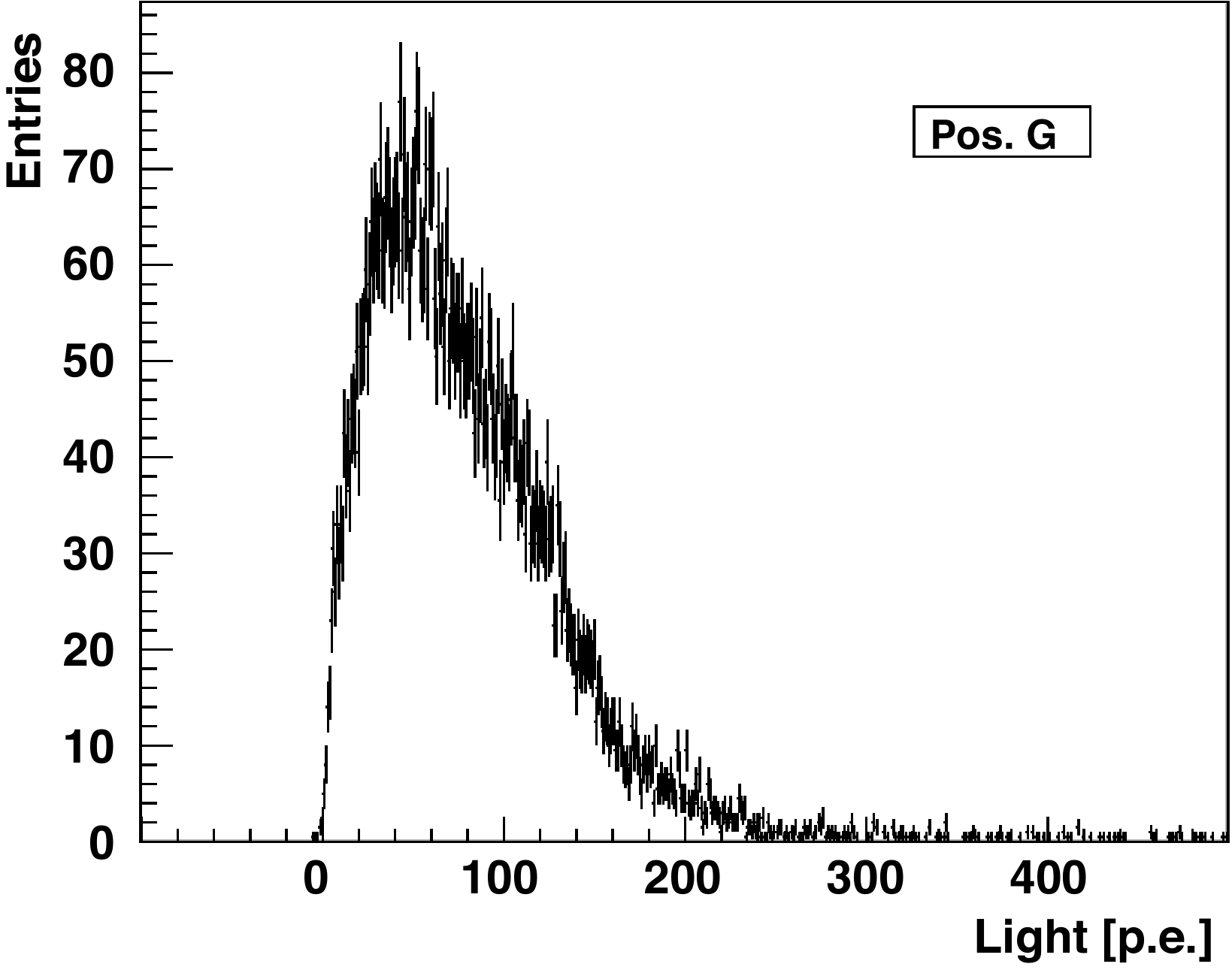}
\caption{Pulse area (p.e.) distribution of electron recoil events 
close to the reconstruction threshold.} 
\label{fig:LYandlowE}
\end{center}
\end{figure}

Since our MC simulation reproduces well our data, we can estimate 
the absolute light yield per unit energy for electron recoils from it. We find that
the average light yield varies between 0.3-0.5~p.e./keVee with the temporary 7~PMTs configuration. 
This is roughly half of the yield that would be obtained with a 
completed detector with 14 PMTs.  Since the light yield for nuclear recoils is lower due to quenching, and 
is typically 25\% that for electrons in the few 10 keV range \cite{Gastler:2010sc}, we expect from these first measurements to
detect 30 keVr nuclear recoils with an average signal of 6 photoelectrons and a resolution of about 40\%.  

For completeness, the 
energy distribution for events passing the trigger conditions of at least two coincident photoelectrons 
within  10\,ns is plotted in fig. \,\ref{fig:LYandlowE}. This also shows that reconstructing events down to a 
few keV is feasible. Further work will be performed to improve the trigger at low energy (this was not
achievable during our first test on surface due to the very high count rate from ambient backgrounds) once
the detector is underground, in a low background environmental condition.

\subsection{Completion of the light readout with 14 PMTs}

\begin{figure}[htb]
\centering
\includegraphics[height=3.7cm]{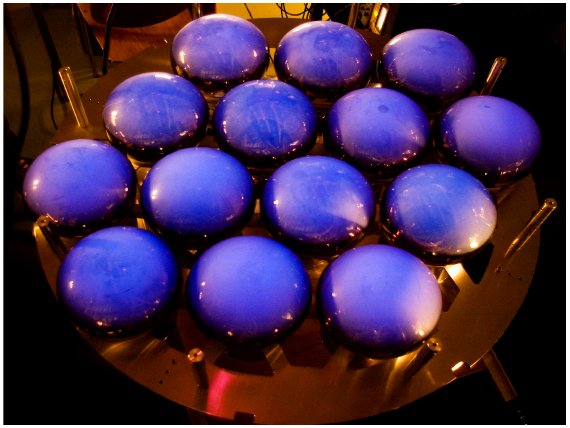}
\includegraphics[height=3.7cm]{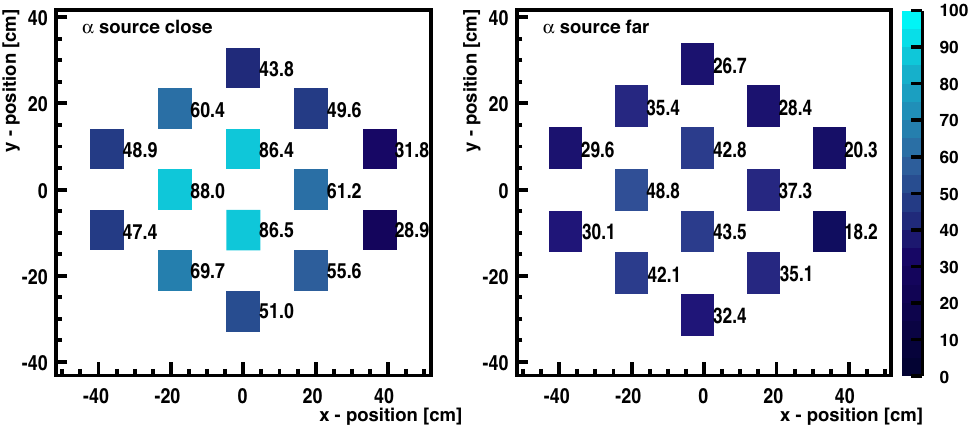}
\caption{Array of 14 cryogenic PMTs under UV illumination. The two plots on the right 
show the detected light intensity (in p.e.)  on the PMTs from the near and far $\alpha$-source in gaseous argon.}
\label{fig:PMT14}
\end{figure}

The light detection system with 14 PMTs was completed recently. Figure \,\ref{fig:PMT14} (left) shows the 
arrangement under illumination with a UV hand held light source, to demonstrate the wavelength shifting behavior of 
the coatings. 
Figure \,\ref{fig:PMT14} (right) displays the results from a test in gaseous argon 
with a vertically movable 5.3 MeV Am $\alpha$-source. The light distributions are 
as expected, showing the homogeneous response of the 14 PMTs when the source 
is not too close to their surfaces.

\section{Conclusions}

Summarizing, we have for the first time experimentally demonstrated that the detection of low energy depositions as those expected from 
nuclear recoils is feasible in the ArDM-1t detector, and we have measured the time and amplitudes distribution of argon luminescence 
during a month long calibration run. The long term purity of the cryogenic system has been established from 
the continuous monitoring of the slow component of the LAr scintillation light. With half of the foreseen light collection 
system installed, a light yield averaged over the active volume of 0.4~photoelectron/keV for electron 
recoils (at zero electric field) has been measured. Most important, we have reached a good understanding of the light 
propagation and detection in our detector, fully implemented in a detailed Monte Carlo simulation.

\acknowledgments

This work was supported by ETH Z\"urich,  the University of Z\"urich and the Swiss National Science Foundation (SNF). We are grateful
to CERN for their hospitality where the tests on surface could be performed, 
and thank the thin film and surface workshop of A.~Braem at CERN for the coating of the PMTs windows.

\end{document}